\documentclass[aps,prl,twocolumn,superscriptaddress]{revtex4-2}
\usepackage{amssymb}
\usepackage{physics}
\usepackage{graphicx}
\usepackage{amsmath}
\usepackage{amsthm}
\usepackage{amsfonts}
\usepackage[T1]{fontenc}
\usepackage[latin9]{inputenc}
\usepackage{array}
\usepackage{multirow}
\usepackage{color}
\usepackage{esint}
\usepackage{bm}
\usepackage{color}
\usepackage{bbm}
\usepackage{hyperref}
\usepackage{babel}
\usepackage{titlesec}
\newcommand{\PT}{\mathcal{PT}}

\newcommand{\SchEq}{Schr\"{o}dinger equation}
\newcommand{\bnj}{\bar{n}_j}
\newcommand{\modelname}{$t_1$-$t_2$}
\newcommand{\lettersection}[1]{\emph{#1}.---}

\newcommand{\Emax}{E_\text{amp}}
\usepackage{hyperref}
\hypersetup
{	colorlinks,%
	citecolor=green,%
	linkcolor=red,%
	urlcolor=blue%
}
\allowdisplaybreaks[4]
\begin{document}

\global\long\def\id{\mathbbm{1}}
\global\long\def\ui{\mathbbm{i}}
\global\long\def\ud{\mathrm{d}}

\title{A new class of exact mobility edges in non-Hermitian quasiperiodic models}

\author{Xu Xia}
\affiliation{Chern Institute of Mathematics and LPMC, Nankai University, Tianjin 300071, China}
\author{Ke Huang}
\affiliation{School of Physics and Technology, Wuhan University, Wuhan 430072, China}
\author{Shubo Wang}
\affiliation{Department of Physics, City University of Hong Kong, Kowloon, Hong Kong SAR}
\affiliation{City University of Hong Kong Shenzhen Research Institute, Shenzhen 518057, Guangdong, China}
\author{Xiao Li}
\email{xiao.li@cityu.edu.hk}
\affiliation{Department of Physics, City University of Hong Kong, Kowloon, Hong Kong SAR}
\affiliation{City University of Hong Kong Shenzhen Research Institute, Shenzhen 518057, Guangdong, China}

\begin{abstract}
Quantum localization in 1D non-Hermitian systems, especially the search for exact single-particle mobility edges, has attracted considerable interest recently. 
While much progress has been made, the available methods to determine the ME in such models are still limited. 
In this work we use a new method to find a new class of exact mobility edges in 1D non-Hermitian quasiperiodic models with parity-time ($\PT$) symmetry. 
We illustrate our method by studying a specific model. 
We first use our method to determine the energy-dependent mobility edge as well as the spectrum for localized eigenstates in this model. 
We then demonstrate that the metal-insulator transition must occur simultaneously with the spontaneous $\PT$-symmetry breaking transition in this model. 
Finally, we propose an experimental protocol based on a 1D photonic lattice to distinguish the extended and localized single-particle states in our model. 
The results in our work can be applied to studying other non-Hermitian quasiperiodic models. 
\end{abstract}

\maketitle

\lettersection{Introduction} 
Quantum localization in disordered media has been a central topic in condensed matter physics since the seminal work by P.~W.~Anderson in 1958~\cite{Anderson1958}. 
In particular, while an infinitesimal amount of disorder will localize all eigenstates in 1D and 2D systems, the full localization transition in 3D systems will only occur at a finite disorder strength~\cite{Abrahams1979_PRL,Lee1985_RMP,Evers2008_RMP}. 
At weaker disorders, however, localized and extended eigenstates in 3D systems can coexist in the energy spectrum, leading to the appearance of a mobility edge (ME). 

Recently, quasiperiodic systems have emerged as a viable alternative platform to study quantum localization in the experiment, partly because they are much easier to realize than those with random disorders. 
Importantly, they have been widely used in the experimental investigation of many-body localization (MBL) in 1D and 2D systems~\cite{Schreiber2015_Science,Bordia2016_PRL,Choi2016_Science,Lueschen2017_PRX,Lueschen2017_PRL,Bordia2017_PRX,Abanin2019_RMP}. 
Moreover, the existence of ME in 1D quasiperiodic systems has also been studied extensively in theory~\cite{Soukoulis1982_PRL,DasSarma1986_PRL,DasSarma1988_PRL,Thouless1988_PRL,DasSarma1990_PRB,Biddle2009_PRA,Biddle2010_PRL,Biddle2011_PRB,Ganeshan2015_PRL,Deng2019_PRL,Li2020_PRB,Wang2020_PRL,Xu2020_NJP,Roy2021_PRL}. 
Such efforts culminated in the recent experimental observation of ME in various 1D systems~\cite{Li2017_PRB,Lueschen2018_PRL,An2018_PRX,Kohlert2019_PRL,Goblot2020_NatPhys,An2021_PRL}. 

Meanwhile, Anderson localization in non-Hermitian systems~\cite{Hatano1996_PRL,Hatano1997_PRB,Hatano1998_PRB,Feinberg1999_PRE,Jazaeri2001_PRE,Molinari2009_JPA,Jovic2012_OL,Yuce2014_PLA,Liang2014_PRA,MejiaCortes2015_PRA,Amir2016_PRE,Harter2016_PRA,Longhi2019_PRL,Tzortzakakis2020_PRB,Huang2020_PRB,Zeng2020_PRB,Xu2021_PRA}, especially the existence of ME in such systems~\cite{Zeng2017_PRA,Longhi2019_PRB,LiuY2020_PRB,LiuT2020_PRB,Zeng2020_PRR,LiuY2021_PRB,Longhi2021_PRB,LiuY2020_arXiv}, have attracted considerable interest recently. 
In particular, much attention has been devoted to systems with the parity-time symmetry ($\PT$ symmetry). 
This symmetry guarantees that the energy spectrum is entirely real when the non-Hermitian parameter $\lambda$ is below a critical value $\lambda_c$;
only when $\lambda>\lambda_c$ complex energies emerge in the spectrum~\cite{Bender1998_PRL,Bender2007_Review,ElGanainy2018_NatPhys}. 
In addition, several properties unique to non-Hermitian systems have also been identified, such as the non-Hermitian skin effects and the existence of exceptional points. 
However, several critical open questions still remain open in this field. 
Notably, most existing work determines the exact ME in a non-Hermitian model using self-duality relations, similar to what has been done in their Hermitian counterparts. 
As a result, when we turn off the non-Hermitian parameter, the model is still known to have an exact ME. 
Can we develop a new method to determine the expression of ME in order to circumvent this limit? 
Crucially, is it possible that a non-Hermitian quasiperiodic model carries an exact ME while its Hermitian counterpart is not known to have one? 
Another critical question is that the existence of ME in a non-Hermitian system has not been experimentally established yet. 
This is partially due to the fact that models with exact MEs are difficult to construct, and thus they often involve a complicated hopping structure or fine-tuned onsite potentials. 
Thus, a non-Hermitian model that can be easily implemented in the experiments is highly desirable.


In this work we address the above questions by studying the localization properties of a 1D non-Hermitian quasiperiodic model with $\PT$ symmetry [see Eq.~\eqref{Eq:TheModel}], which reduces to the Hermitian {\modelname} model~\cite{Biddle2009_PRA,Biddle2010_PRL,Biddle2011_PRB} when the non-Hermitian parameter is turned off. 
We show that the ME in this model can be determined analytically by the Sarnak method~\cite{Sarnak1982}. 
This result is remarkable, because the exact ME in the Hermitian {\modelname} model is not yet known. 
Moreover, while the spectrum of the Hermitian {\modelname} model has a fractal structure, the spectrum of our model is dense.   
In fact, the Sarnak method can help us analytically determine the entire spectrum of localized states. 
Thus our model is fundamentally different from its Hermitian counterpart. 
Additionally, we demonstrate that the metal-insulator transition in this model must occur simultaneously with the spontaneous $\PT$ symmetry breaking transition.  
Further, we demonstrate that the ME only exists for a finite range of potential strengths $V_{c1} \leq V \leq V_{c2}$, and determine $V_{c1}$ and $V_{c2}$ exactly. 
Finally, we propose an experimental protocol based on a 1D photonic lattice to distinguish extended and localized states in this model.

\lettersection{Model}
To begin with, consider the following non-Hermitian quasiperiodic model, 
\begin{align}
H = \sum_{ j}\qty(t_1c^{\dagger}_{j}c_{j+1}+t_2c^{\dagger}_{j}c_{j+2} + \text{h.c.})
+\sum_j V_j n_{j} . \label{Eq:TheModel}
\end{align}
In the above equation $c_{j}$ annihilates a fermion on site $j$, and $n_{j}=c^{\dagger}_{j}c_{j}$ counts the particle number on site $j$.
For convenience, we set the hopping strength $t_1=1$ as the unit of energy. 
In addition, we only consider the cases with $t_2>0$, as the $t_2<0$ can be easily reduced to the $t_2>0$ case.  
The potential energy in Eq.~\eqref{Eq:TheModel} is given by $V_j=Ve^{2\pi i (\phi+j\alpha)}$ with $V>0$. 
Without loss of generality, we will set $\phi=0$. 
Finally, we take $\alpha=(\sqrt{5}-1)/2$, which can be approximated by Fibonacci numbers $F_{n}$~\cite{Kohmoto1983_PRL,Wang2016_EPJ}: $\alpha=\lim_{n \rightarrow \infty}{F_{n-1}}/{F_{n}}$. 
Specifically, in our simulations we choose a specific integer $n$ so that the system size is $L=F_{n}$ and $\alpha=F_{n-1}/F_{n}$. 
This choice ensures the $\PT$ symmetry in our model. 

\lettersection{The localization transition and ME} 
As one of the key results in this Letter, we find that the model in Eq.~\eqref{Eq:TheModel} possesses an energy-dependent ME, given exactly by the following analytical expression,
\begin{align}
V=\frac{1}{4}\abs{1+\sqrt{\Delta}+\sqrt{\left(1+\sqrt{\Delta}\right)^2-16 t_2^2}},
\label{Eq:ME}
\end{align}
where $\Delta=1+4 t_2 E+8 t_2^{2}$, and $E\in[2t_2-2,2t_2+2]$ specifies the range of energies at which an ME can exist~\footnote{Note that the energy spectrum is guaranteed to be real at the ME.}. 
As we show below, this ME marks the simultaneous metal-insulator transition and the spontaneous $\PT$-symmetry breaking transition in this model. 
In fact, we can use the Sarnak method~\cite{Sarnak1982} to derive an analytical condition for the spectrum of localized states in this model, given by 
\begin{align}
    \log|V|=G(E),  \label{Eq:LocalizationSpectrum} 
\end{align}
where $G(E)$ is defined as~\cite{SM} 
\begin{align}
    G(E)=\frac1{2\pi}\int_{0}^{2\pi}
    \log\abs\Big{E-2\cos\theta-2t_2\cos2\theta}d\theta.
\end{align}
The ME condition in Eq.~\eqref{Eq:ME} can be viewed as a special case of Eq.~\eqref{Eq:LocalizationSpectrum} when $E\in[2t_2-2,2t_2+2]$.

\begin{figure}[t]
\includegraphics[width=0.48\textwidth]{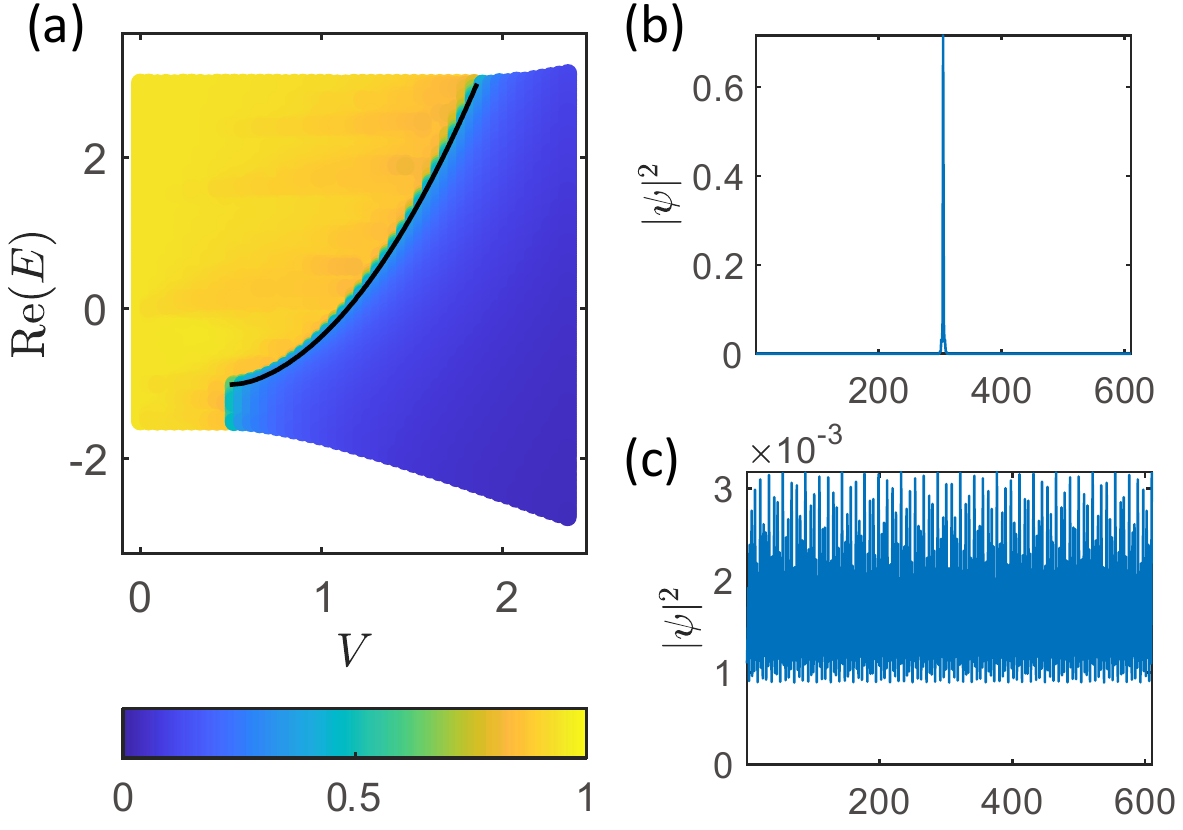}
\caption{\label{Fig:ME}
(a) The fractal dimension $\Gamma$ as a function of $\Re(E)$ and $V$ in a lattice with size $L=F_{14}=610$. 
The black line represents the ME condition in Eq.~\eqref{Eq:ME}. 
(b) and (c) plot the wave function for the two states at $V = (V_{c1}+V_{c2})/2$, which has the largest and smallest $\Re(E)$, respectively. 
Here we choose $t_2=1/2$ for all three figures. 
}
\end{figure}

One convenient tool to identify localized states is the inverse participation ratio (IPR), defined as ${\rm IPR}(m) = \sum_{j} |\psi_{m,j}|^4$~\cite{Evers2008_RMP,Li2017_PRB}, where $m$ labels the eigenstates and $j$ labels lattice sites. 
Based on this, we can further introduce the fractal dimension of the wave function, $\Gamma=-\lim_{L\rightarrow\infty}\frac{\ln({\rm IPR})}{\ln L}$. 
One can show that for extended states $\Gamma\to 1$ while for localized states $\Gamma\to 0$. 
In Fig.~\ref{Fig:ME}(a) we plot the fractal dimension $\Gamma$ of each eigenstate as a function of $\Re(E)$ and $V$. 
In addition, the black line represents the ME condition in Eq.~\eqref{Eq:ME}.  
As expected, $\Gamma$ approaches zero and one for energies on opposite sides of the black line, respectively. 
This can be further confirmed by the spatial density profile of the respective eigenstates, see Fig.~\ref{Fig:ME}~(b)-(c). 
In other words, a given eigenstate is localized or extended depends on whether its eigenvalue satisfies $\log\abs{V}\leq G(\Re(E))$ or $\log\abs{V} > G(\Re(E))$~\cite{SM}. 

\begin{figure}[t]
\includegraphics[width=0.48\textwidth]{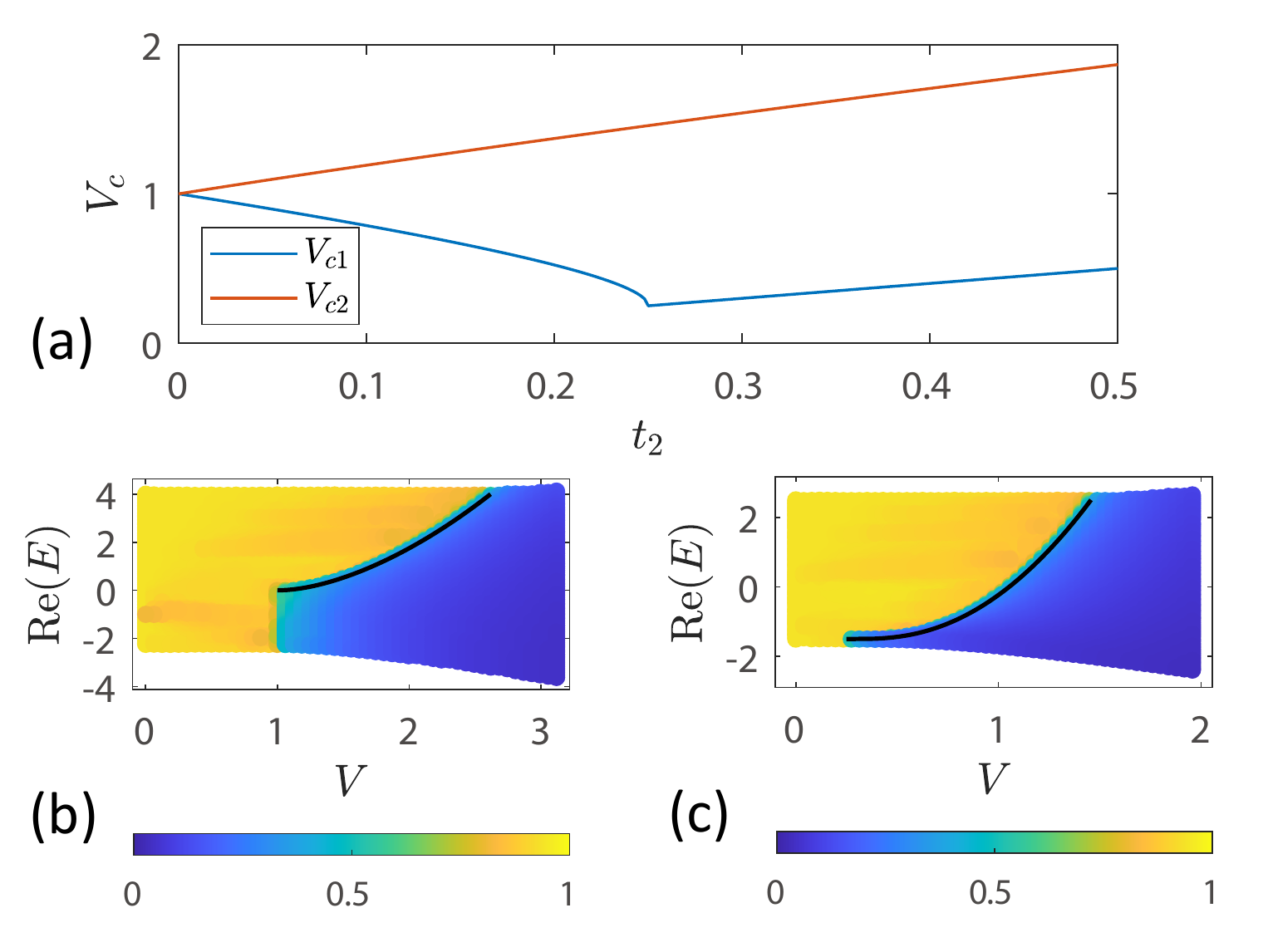}
\caption{\label{Fig:PhaseDiagram}
(a) The boundaries of the intermediate phase as a function of $t_2$. 
$V_{c1}$ and $V_{c2}$ denote the critical $V$ at which the intermediate phase starts and ends for a specific $t_2$, see Eq.~\eqref{Eq:Vc1&Vc2}. 
(b)-(c) Fractal dimension $\Gamma$ of each eigenstate for $t_2=1$ and $t_2=1/4$, respectively. 
The black lines represent the ME described by Eq.~\eqref{Eq:ME}. 
Here the system size is $L=610$.  
}
\end{figure}

We can thus identify three distinct regimes in Fig.~\ref{Fig:ME}(a): 
for $V<V_{c1}$ ($V>V_{c2}$), the energy spectrum only contains extended (localized) eigenstates, while for $V_{c1}\leq V \leq V_{c2}$, an energy-dependent ME emerges. 
We will thus denote the regime $V_{c1}\leq V \leq V_{c2}$ as the intermediate phase, since both extended and localized states exist in the spectrum. 
More importantly, we find that an intermediate phase always exists when $t_2\neq 0$, and that the exact expressions for $V_{c1}$ and $V_{c2}$ are given by~\cite{SM} 
\begin{align}
    \begin{aligned}
         V_{c1}&=
         \begin{cases}
            t_2, & t_2 \geq 1/4 \\
            \frac{1}{2}\qty(\sqrt{1-4t_2} + 1-2t_2), & 0 \leq t_2 <1/4. 
         \end{cases}
         ,\\
         V_{c2}&=\frac12\qty(\sqrt{1 + 4t_2} + 1 + 2t_2),
    \end{aligned} 
    \label{Eq:Vc1&Vc2}
\end{align}
which are plotted in Fig.~\ref{Fig:PhaseDiagram}(a). 

Interestingly, Fig.~\ref{Fig:PhaseDiagram}(a) shows a curious cusp in $V_{c1}$ at $t_2=1/4$, which implies that $t_2\leq 1/4$ and $t_2>1/4$ are two different regimes. 
This conjecture is confirmed in Fig.~\ref{Fig:PhaseDiagram}~(b)-(c), where we plot the ME for $t_2=1$ and $t_2=1/4$, respectively. 
We find that when $t_2>1/4$ [Fig.~\ref{Fig:PhaseDiagram}(b)], the number of localized states suddenly becomes finite as $V$ crosses $V_{c1}$. 
In contrast, when $t_2\leq1/4$ [Fig.~\ref{Fig:PhaseDiagram}(c)], the number of the localized states increases continuously from zero as $V$ crosses $V_{c1}$. 
Therefore, we conclude that the structure of the ME is qualitatively different when $t_2\leq 1/4$ and $t_2>1/4$.

\begin{figure}[t]
\includegraphics[width=0.48\textwidth]{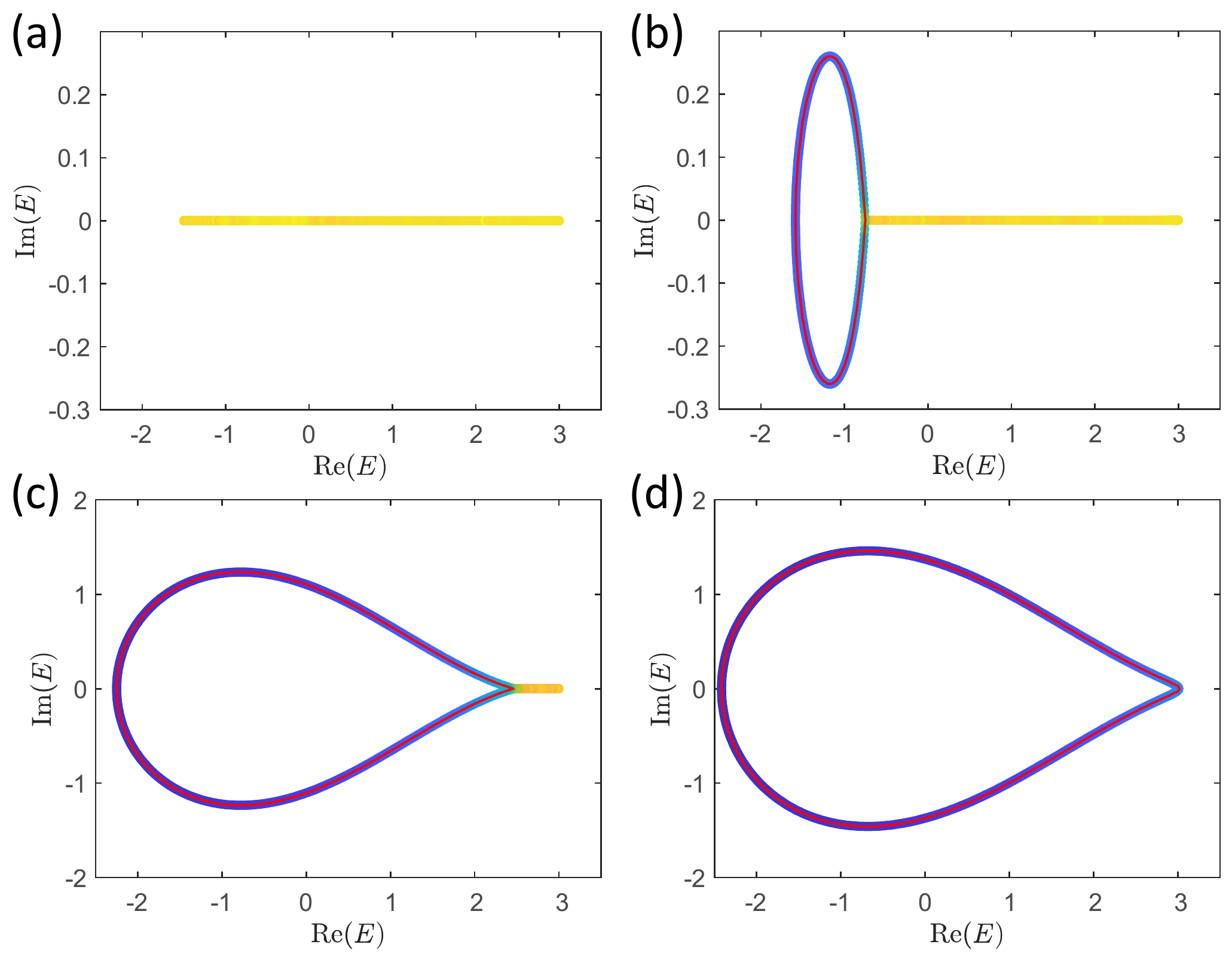}
\caption{\label{Fig:Spectrum}
The complex energy spectrum for (a) $V=V_{c1}-0.1$, (b) $V_{c1}+0.3$, (c) $V_{c2}-0.1$, and (d) $V_{c2}+0.1$. 
The color of the energy spectrum represents the fractal dimension $\Gamma$ of the eigenstates using the same color scale as that in Fig.~\ref{Fig:ME}.
In addition, the red lines in (b)-(d) map out the spectrum of localized states for the corresponding $V$ [see Eq.~\eqref{Eq:LocalizationSpectrum}]. 
Here we fixed the system size to be $L=610$, and keep $t_2=1/2$.  
}
\end{figure}

\lettersection{The $\PT$-symmetry breaking transition}
Apart from the metal-insulator transition described above, another interesting property of a $\PT$-symmetric non-Hermitian model is that this symmetry can be spontaneously broken when the non-Hermitian parameter $V$ exceeds a critical value. 
Moreover, it is known that this phase transition is accompanied by the transition from an entirely real spectrum to a complex one~\cite{Bender1998_PRL,Bender2007_Review,ElGanainy2018_NatPhys}. 
To demonstrate this property in our model, we keep $t_2=1/2$ and plot in Fig.~\ref{Fig:Spectrum} the spectrum for $V$ around $V_{c1}$ and $V_{c2}$, respectively. 
The results show that the analytical condition in Eq.~\eqref{Eq:LocalizationSpectrum} (shown as red lines in Fig.~\ref{Fig:Spectrum}) correctly captures the spectrum of localized states. 
In addition, we can observe two different transitions. 
First, as $V$ increases beyond $V_{c1}$, complex energies start to emerge from a purely real spectrum, which is accompanied by the appearance of localized states. 
Second, as $V$ further increases beyond $V_{c2}$, the spectrum turns into a purely complex one and no extended states exist anymore.
It is well known that in a $\PT$ symmetric model the spontaneous $\mathcal{PT}$ symmetry breaking underlies the real-complex transition of the spectrum. 
What is particularly interesting about our model is that the seemingly unrelated metal-insulator transition occurs simultaneously with the spontaneous $\PT$ symmetry breaking transition.
In fact, we can prove this property rigorously, see Ref.~\cite{SM}.

\begin{figure*}[t]
\includegraphics[width=1\textwidth]{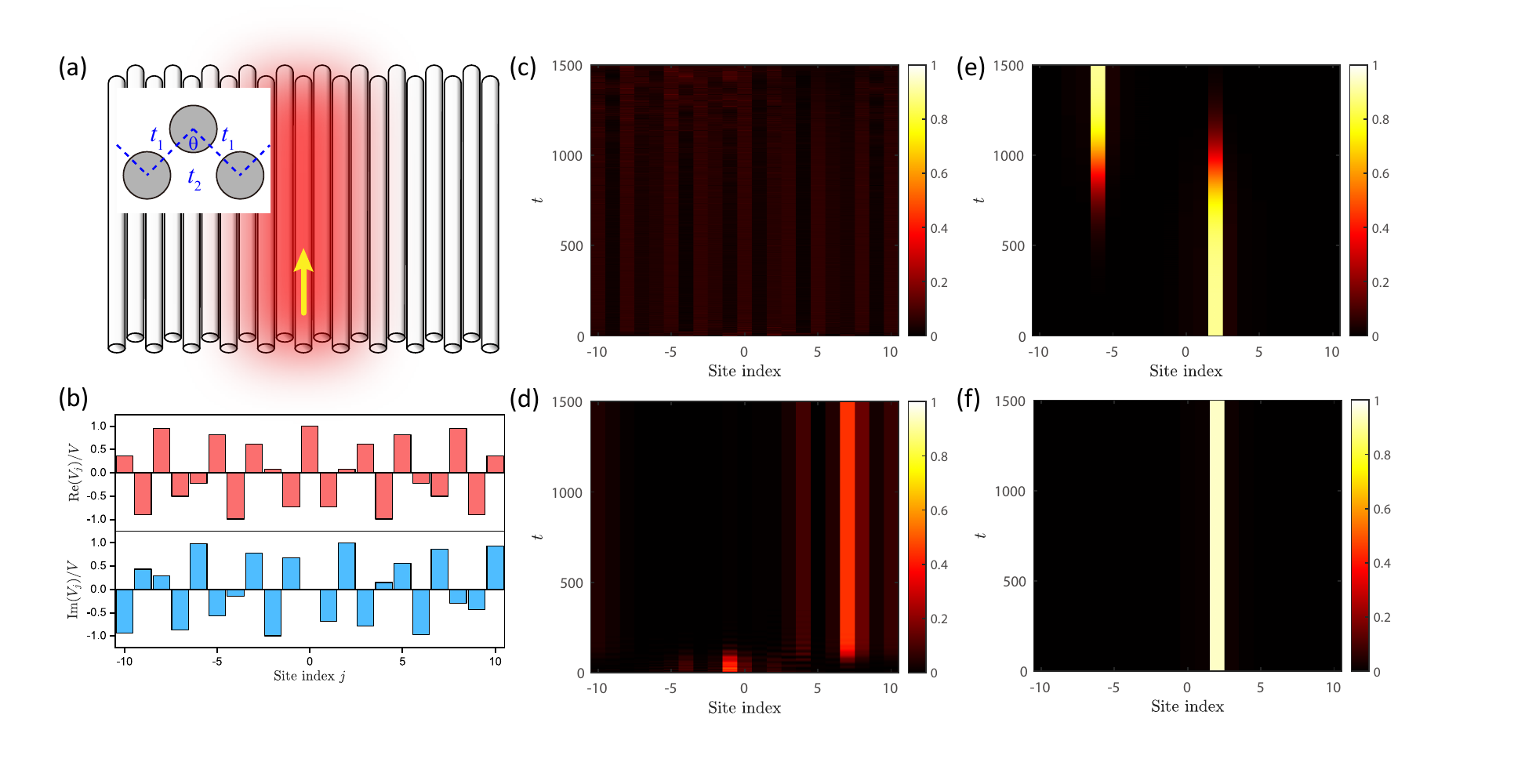}
\caption{\label{Fig:Exp}
A realistic experimental realization of the non-Hermitian $t_1$-$t_2$ model in Eq.~\eqref{Eq:TheModel}. 
(a) The schematic setup of a coupled waveguide system consisting of $L=21$ waveguides. 
The inset illustrates the coupling between them. 
Note that the ratio $t_2/t_1$ can be tuned by varying the angle $\theta$. 
The arrow indicates that the initial excitation occurs in the $j=0$ waveguide. 
(b) A plot of the onsite potential $V_j$ in this coupled waveguide system. 
The red and blue bars illustrate the real and imaginary part of the potential, respectively. 
(c)-(f) show the quench dynamics starting from an initial excitation in the $j=0$ waveguide for $V=V_{c1}-0.1,~V_{c1}+0.3,~V_{c2}+1,~V_{c2}+2$, respectively. 
In this plot we choose $\alpha=13/21$ and $t_2/t_1=1/2$. 
The color bars in (c)-(f) plot the $\bnj(t)$ defined in Eq.~\eqref{Eq:bnj}. 
The unit of time is $\tau=\hbar/t_1$. 
The short-time behavior of $\bnj(t)$ can be found in ~\cite{SM}. 
}
\end{figure*}

{\lettersection{Experimental realizations}}
We now present a realistic experimental realization of the non-Hermitian $t_1$-$t_2$ model in Eq.~\eqref{Eq:TheModel} using a photonic lattice. 
Such photonic lattices have been routinely used to demonstrate Anderson localization of light~\cite{Schwartz2007_Nature,Lahini2008_PRL}. 
A schematic setup of our proposal is shown in Fig.~\ref{Fig:Exp}(a). 
It is known that in the paraxial limit the propagation of classical light in a waveguide can be captured by a form of Maxwell equation that formally resembles the {\SchEq} in quantum mechanics~\cite{Ozawa2019_Review,SM}. 
If we further consider the limit in which the light is strongly confined by the waveguides, one can adopt the tight-binding approximation, and cast the continuum wave equation in the following form~\cite{Ozawa2019_Review}, 
\begin{align}
    i\dv{\psi_j}{z} = \kappa_j \psi_j + \sum_{l\neq j} J_{j,l} \psi_{l}. \label{Eq:TB_Model}
\end{align}
Here the wave vector $\kappa_j$ is controlled by the refractive index contrast of the $j$th waveguide and the background medium, while the tunneling parameters $J_{j,l}$ are determined by the overlap between the evanescent tails of the eigenmodes in the $j$th and $l$th waveguides~\cite{Ozawa2019_Review}.  

Our model in Eq.~\eqref{Eq:TheModel} can be realized in such a coupled waveguide system where the refractive index in the $j$th waveguide plays the role of potential $V_j$ and the temporal coordinate $t$ is replaced by the spatial coordinate $z$. 
In this work, we choose a system of $L=21$ coupled waveguides, see Fig.~\ref{Fig:Exp}(a). 
In particular, it is possible to engineer the refractive indices of the waveguides so that their real and imaginary parts resemble the complex potential as plotted in Fig.~\ref{Fig:Exp}(b). 
We further set $J_{j,j+1} = J_{j,j-1}= t_1$, $J_{j,j+2}=J_{j,j-2} = t_2$, and all other $J_{j,l}=0$. 
Further, the waveguides are arranged in a zigzag shape, so that the nearest neighbor coupling $t_1$ is larger than the next nearest neighbor coupling $t_2$. 
In this geometry the ratio $t_2/t_1$ can be tuned by the angle $\theta$ of the zigzag chain, see the inset of Fig.~\ref{Fig:Exp}(a). 
Finally, periodic boundary conditions are preferred in the setup. 

The localization property of this model can be probed by studying the light propagation in this coupled waveguide system. 
Here we choose to excite the waveguide at $j=0$ at $t=0$, and study how the light spreads out during the propagation. 
Effectively, we are evaluating 
\begin{equation}
    \ket{\psi(t)} = e^{-iHt}\ket{\psi_0} 
    = \sum_{j}e^{-iE_jt}c_j\ket{E_j}, 
\end{equation}
where $\ket{E_j}$ is the $j$th eigenstate of the Hamiltonian $H$ in Eq.~\eqref{Eq:TheModel} with an energy $E_j$, and $\qty{c_j}$ are the superposition coefficients. 
The spatial extent of the time evolved state $\ket{\psi(t)}$ can be quantified by
\begin{align}
    \bnj(t) \equiv \dfrac{\abs{\bra{w_j}\ket{\psi(t)}}^2}{\bra{\psi(t)}\ket{\psi(t)}}, \label{Eq:bnj}
\end{align}
where $\ket{w_j}$ denotes the Wannier function localized within the $j$th waveguide. 

We first consider the $V<V_{c1}$ regime, when all eigenstates in the system are extended. 
Consequently, we expect that almost all $\bnj(t)$ are nonzero at late times.
In addition, because the spectrum is completely real, all the phase factors $e^{-iE_jt}$ satisfy $\abs{e^{-iE_jt}}=1$ at all times.  
As a result, all eigenstates will continue to contribute to the dynamics even when $t$ is large.  
Our expectations are verified by the results in Fig.~\ref{Fig:Exp}(c), where we numerically plot $\bnj(t)$ when $V = V_{c1} - 0.1$. 
In particular, we find that within a short time the initial excitation spreads out to other waveguides, and $\bnj(t)$ is almost evenly distributed among all waveguides.  

In contrast, when the energy spectrum is complex, the time evolution operator $e^{-iHt}$ is dominated by the eigenstate whose energy eigenvalue has the largest imaginary part.  
For convenience, we denote this special eigenvalue as $\Emax$ and the corresponding eigenstate as $\ket{\Emax}$. 
In order to avoid numerical errors induced by the exponential amplifications in the presence of a complex spectrum, we further replace the original Hamiltonian $H$ by $H' = H - i\gamma$ in our simulations, where $\gamma \equiv \Im(\Emax)>0$.
As a result, the state $\ket{\Emax}$ still dominates the quench dynamics, but its amplitude is preserved throughout the dynamics. 
In contrast, the amplitude of all the other eigenstates decays exponentially. 
Furthermore, since in this model all states with a complex energy eigenvalue are localized, we anticipate that the final state will be localized whenever the spectrum contains complex energies. 
In Fig.~\ref{Fig:Exp}(d) we plot $\bnj(t)$ for $V=V_{c1}+0.3$, when the system is in the intermediate phase. 
We indeed find that the final state is localized. 
However, in contrast to the quench dynamics in a Hermitian system, the final state is not localized on the original waveguide at $j=0$, but collapses into the waveguide at $j=7$. 
Moreover, we find a curious `switching process' during the dynamics. 
Specifically, the initial excitation in the $j=0$ waveguide almost instantly switches to a signal peaked at the $j=-1$ waveguide~\cite{SM}. 
At around $t=100$, this signal switches again to one localized in the $j=7$ waveguide. 
During the entire quench dynamics, the maximum magnitude of $\bnj(t)$ reaches about $0.4$. 
We also find that the localization length of the final steady state is still quite large, as weak signals with $\bnj(t)\sim 0.2$ can still be seen in the neighboring waveguides at $j=4$ and $j=10$. 
The above observations show that localization in a non-Hermitian system is qualitatively different from that in Hermitian systems. 
In particular, the switching behavior can never occur in a Hermitian system. 


In addition, in Fig.~\ref{Fig:Exp} (e)-(f) we plot $\bnj(t)$ for two different $V>V_{c2}$, when the system is in the localized phase. 
We find that the qualitative features of Fig.~\ref{Fig:Exp}(d), especially the switching behavior, are preserved. 
For example, in Fig.~\ref{Fig:Exp}(e) we find that the initial excitation in the $j=0$ waveguide quickly gives way to an excitation confined in the $j=2$ waveguide, before eventually collapses into the waveguide at $j=-6$.  
In comparison, in Fig.~\ref{Fig:Exp}(f) we find that the initial excitation in the $j=0$ waveguide quickly collapses into the waveguide at $j=2$ and no additional switching happens afterwards. 
The main differences between the localized regime and the intermediate regime seem to be quantitative. 
For example, the localization length of the final steady state is now reduced to just one lattice site. 
Moreover, the peak value of $\bnj(t)$ now reaches about $0.8$ for Fig.~\ref{Fig:Exp}(e) and about $0.9$ for Fig.~\ref{Fig:Exp}(f), respectively. 
It turns out that the curious switching behavior of $\bnj(t)$ found in Fig.~\ref{Fig:Exp}~(d)-(f) arise because there exist several eigenstates whose eigenvalues have similar imaginary parts. 
The switching is a result of the competitions between these eigenstates~\cite{SM}.

\lettersection{Discussion and Outlook} 
Our work represents one of the first examples where the ME in the non-Hermitian quasiperiodic model cannot be directly inferred from its Hermitian counterpart. 
Indeed, while the exact ME in the Hermitian {\modelname} model is not yet known, we are able to determine the exact ME in our model. 
In addition, the method developed in this work is very general and can be applied to a wide class of quasiperiodic models. 
For example, an exact ME can still be obtained when $t_2$ is complex or when more remote hopping terms are included~\cite{SM}. 
Our work thus not only proposes a realistic experimental scheme to demonstrate ME in a non-Hermitian quasiperiodic model, but also presents a general framework to study other 1D non-Hermitian quasiperiodic models. 
One important open question is the effect of interactions on the localization properties of this model~\cite{Hamazaki2019_PRL,Zhai2020_PRB}. 
In particular, it is interesting to understand whether the interplay between interactions and the ME can lead to a many-body intermediate phase~\cite{Hsu2018_PRL,Xu2019_PRR,Wang2021_PRL} in this non-Hermitian system. 

\lettersection{Acknowledgements}
S.W. acknowledges support from the Research Grants Council of the Hong Kong Special Administrative Region, China (Project~Nos.~HKUST~C6013-18G and CityU~11301820). 
X.L. acknowledges support from City University of Hong Kong (Project~No.~9610428), 
the National Natural Science Foundation of China (Grant~No.~11904305), as well as 
the Research Grants Council of the Hong Kong Special Administrative Region, China (Grant~No.~CityU~21304720). 

\bibliographystyle{apsrev4-2}
\bibliography{t1t2Model_v6.bib}


\end{document}


\global\long\def\id{\mathbbm{1}}
\global\long\def\ui{\mathbbm{i}}
\global\long\def\ud{\mathrm{d}}

\title{Supplemental Material for ``A new class of exact mobility edges in non-Hermitian quasiperiodic models''}

\author{Xu Xia}
\affiliation{Chern Institute of Mathematics and LPMC, Nankai University, Tianjin 300071, China}
\author{Ke Huang}
\affiliation{School of Physics and Technology, Wuhan University, Wuhan 430072, China}
\author{Shubo Wang}
\affiliation{Department of Physics, City University of Hong Kong, Kowloon, Hong Kong SAR}
\affiliation{City University of Hong Kong Shenzhen Research Institute, Shenzhen 518057, Guangdong, China}
\author{Xiao Li}
\email{xiao.li@cityu.edu.hk}
\affiliation{Department of Physics, City University of Hong Kong, Kowloon, Hong Kong SAR}
\affiliation{City University of Hong Kong Shenzhen Research Institute, Shenzhen 518057, Guangdong, China}

\setcounter{equation}{0} 
\setcounter{figure}{0}
\setcounter{table}{0} 
\renewcommand{\theparagraph}{\bf}
\renewcommand{\thefigure}{S\arabic{figure}}
\renewcommand{\theequation}{S\arabic{equation}}
\renewcommand\labelenumi{(\theenumi)}

\maketitle

In this Supplemental Material, we present additional details on the structure of the mobility edge in the non-Hermitian {\modelname} model we studied in the main text. 
We also provide more details on the experimental protocol we presented in the main text. 
Finally, we analyze the cases when the hopping parameters take complex values. 

\section{Additional details of the ME}
In this section we provide additional details of the ME for the model we discussed in the main text, which has the following form, 
\begin{align}
H = \sum_{ j}\qty(t_1c^{\dagger}_{j}c_{j+1}+t_2c^{\dagger}_{j}c_{j+2} + \text{h.c.})
+\sum_j V_j n_{j} . \label{EqSM:TheModel}
\end{align}
Given that the systems with $t_2$ and $-t_2$ are related through following substitution $c_j^\dag\to (-1)^jc_j^\dag$, we will only consider $t_2>0$ in the following discussions. 

\subsection[G(E)]{The expression for $G(E)$}
In order to derive the exact ME in our model, we first introduce the following function,  
\begin{align}
    f(\theta)=\sum_{j \in \mathbb{Z}} \psi_{j} e^{i j \theta} \in L^{2}(\mathbb{T}). \label{EqSM:FourierTransform}
\end{align}
After Fourier transform,  which is multiplying Eq.~\eqref{EqSM:TheModel} by $e^{i j \theta}$ and then summing over $j$, we get the operator 
\begin{align}
V f(\theta+2\pi\alpha)=\qty[E-2\cos\theta-2t_2\cos2\theta]f(\theta),\label{EqSM:EquivalentCondition}
\end{align}
It has been proved that the spectrum of such a system can be captured by a characteristic function defined as
\begin{align}
    G(E)=\frac1{2\pi}\int_{0}^{2\pi}\log\abs\bigg{E-2\cos\theta-2t_2\cos2\theta}d\theta. 
    \label{EqSM:G(E)}
\end{align}
Note that we always have $G(E) \geq G(\Re(E))$, because for any $g(\theta)\in\mathbb{R}$, we have 
\begin{align*}
    \log\abs{E - g(\theta)} \equiv& \log\abs{\Re(E) + i\Im(E) - g(\theta)} \notag\\
   \geq& \log\abs{\Re(E) - g(\theta)}. 
\end{align*}
Therefore, if we take $g(\theta) = 2\cos\theta + 2t_2\cos2\theta$, 
\begin{align*}
    G(E) &= \dfrac{1}{2\pi}\int_0^{2\pi}\log\abs{E-g(\theta)}d\theta \notag\\
    &\geq \dfrac{1}{2\pi}\int_0^{2\pi}\log\abs{\Re(E)-g(\theta)}d\theta \equiv G(\Re(E)). 
\end{align*}

The Sarnak method~\cite{Sarnak1982} then states that: 
\begin{itemize}
\item There is no spectrum within $G(E)<\log{V}$; 
\item $G(E)=\log{V}$ has a dense localized spectrum; 
\item Extended states belong to $\{G(E)>\log{V}\} \cap U_E $, where $U_E$ is the set of energies that satisfy $E = 2\cos(\theta) + 2t_2\cos(2\theta)$ for some $\theta$. 
\end{itemize}
In particular, we can write 
\begin{align}
U_E = 
\begin{cases}
    [-1/(4t_2)-2t_2,2t_2+2], & 	t_2 \geqslant 1/4\\
    [2t_2-2,2t_2+2],  & 0 \leqslant t_2 < 1/4
\end{cases}. 
\end{align}
After some algebra we find that 
\begin{align}\label{G(E)_real}
    G(E)=&\log t_2 + \log\abs{z_1\,\mathrm{sgn}(\mathrm{Re}z_1)+\sqrt{z_1^2-1}} \notag\\
    &+ \log\abs{z_2\,\mathrm{sgn}(\mathrm{Re}z_2)+\sqrt{z_2^2-1}},
\end{align}
where $\mathrm{sgn}(x)$ is the sign of the real number $x$, $\sqrt z$ is the square root of $z$ with non-negative real part. 
In addition, $z_1=\frac1{4t_2}(1+\sqrt{\Delta})$, $z_2=\frac1{4t_2}(1-\sqrt{\Delta})$, with $\Delta=1+4t_2 E+8t_2^2$.
As a result, $G(E)$ always reaches its minimum value at $E=2t_2-2$, which belongs to $U_E$. 
Likewise, $G(E)$ always reaches its maximum value in $U_E$ at $E=2t_2+2$.

\subsection[Vc1 and Vc2]{The expression for $V_{c1}$ and $V_{c2}$}
From the above results, we can determine the structure of the metal-insulator transition in this model, which is accompanied by a $\mathcal{PT}$ symmetry breaking transition in the energy spectrum. 
In particular, we can introduce two critical points to describe this transition: 
\begin{align}
V_{c1} &= e^{G(2t_2-2)} \\
&=\begin{cases}
    t_2, & t_2 \geq 1/4 \\
    \frac{1}{2}\qty(\sqrt{1-4t_2} + 1-2t_2), & 0 \leq t_2 <1/4. 
 \end{cases}, \notag\\
V_{c2} &=e^{G(2t_2+2)}=\frac12\left[2t_2+1+\sqrt{4t_2+1}\right]. 
\end{align}
These two points divide the phase diagram into three regimes. 
For $V$~$<$~$V_{c1}$, all eigenstates are extended. 
Meanwhile, the energy spectrum resides in $U_E$ and is thus entirely real. 
For $V_{c1}$~$<$~$V$~$<$~$V_{c2}$, an ME appears in the energy spectrum, indicating that extended and localized eigenstates coexist in the energy spectrum. 
Meanwhile, we observe that the spectrum contains both real and complex eigenvalues. 
Finally, when $V>V_{c2}$, we have $G(E)<\log|V|$ for all $E\in U_E$, and thus the entire spectrum is localized and complex in general. 

\subsection[PT symmetry breaking transition]{Localization transition and the $\PT$ symmetry breaking transition}

\begin{figure}[!]
\includegraphics[width=\columnwidth]{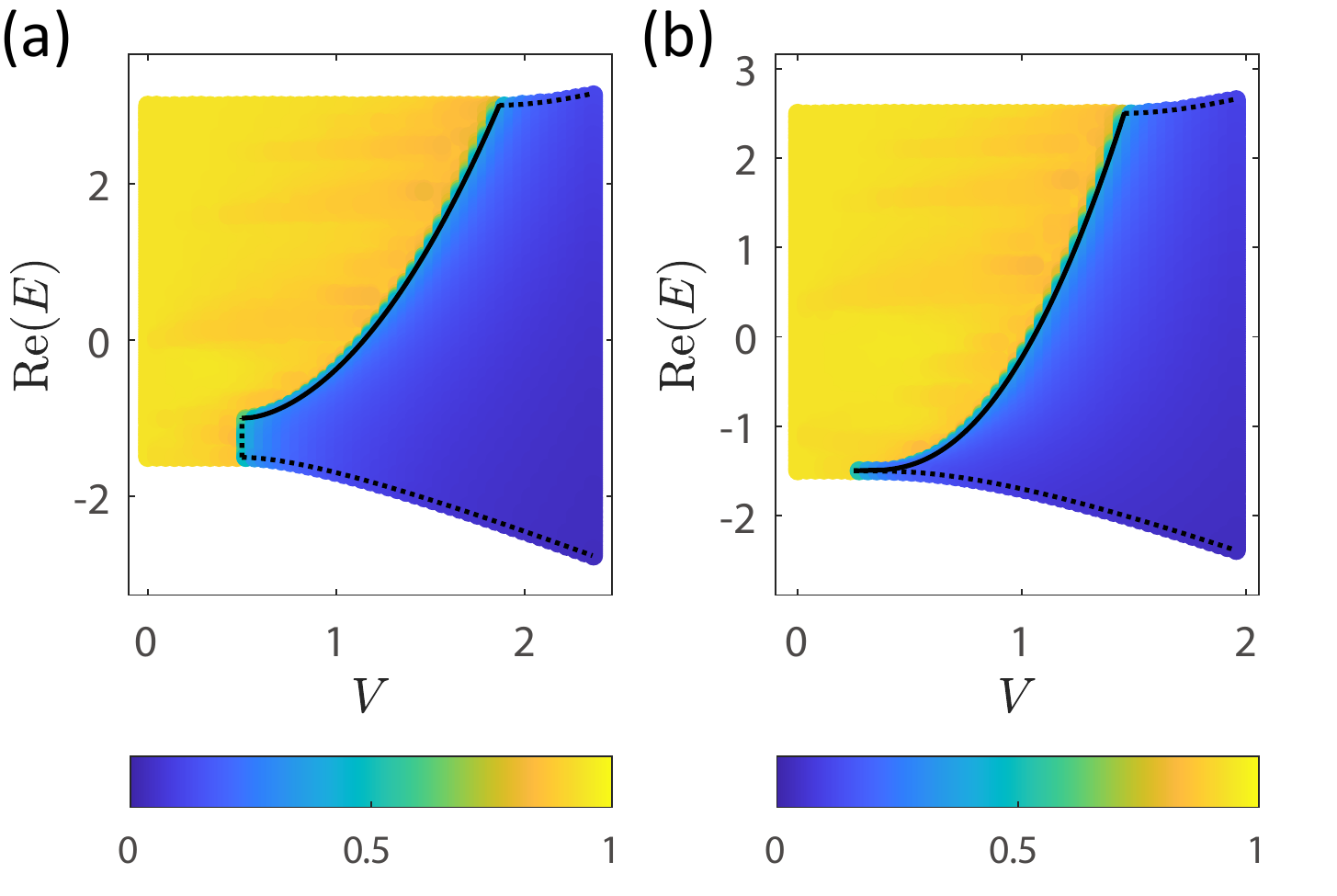}
\caption{\label{FigSM:FD} (a) and (b) Fractal dimension $\Gamma$ of each eigenstate for the model in Eq.~\eqref{EqSM:TheModel} for $t_2=1$ and $t_2=1/4$, respectively. 
The system size is $L=610$. 
The solid black line highlights eigenstates on the ME (which are localized states with a real energy eigenvalue), while the dashed black lines highlight additional localized states with a real energy eigenvalue. 
}
\end{figure}

In fact, we can prove rigorously that in our model [Eq.~\eqref{EqSM:TheModel}] the $\PT$ symmetry breaking transition (which is signaled by the real-complex transition in the energy spectrum) and the localization transition must occur simultaneously. 
First we note that extended spectrum must be real, since $U_E\subset \mathbb{R}$. 
Therefore, we just need to prove localized states must have complex energy eigenvalues. 
However, there is a subtlety here, as few localized states still have real energy eigenvalues. 
In particular, the following localized states are known to have real eigenvalues: 
(1) All eigenstates on the ME (marked by the solid black line in Fig.~\ref{FigSM:FD}); 
(2) Eigenstates at the boundary of the spectrum (marked by dashed black lines in Fig.~\ref{FigSM:FD}). 
However, given that they are all located at the boundary of the spectrum of localized states, they do not affect our claim that the $\PT$ symmetry transition must be accompanied by the metal-insulator transition in this model.  

We now show that apart from those marked by the solid and the dashed black lines in Fig.~\ref{FigSM:FD}, all other localized eigenstates (which occur for $V>V_{c1}$) have complex energy eigenvalues. 
To start, consider the real part of the complex spectrum
\begin{align*}
    \mathcal O=\{\mathrm{Re}(E):G(E)=\log{V}\ \&\ \mathrm{Im}(E)\neq 0\}. 
\end{align*}
As we pointed out below Eq.~\eqref{EqSM:G(E)}, $G(E)\geq G(\Re(E))$ for all $E$. 
As a result, we must have $\mathcal O\subset \mathcal O'$, where 
\begin{align*}
    \mathcal O'=\{x\in\mathbb{R}:G(x)<\log{V}\}.
\end{align*}
We can see clearly that $\mathcal O'$ is a nonempty open set in $\mathbb{R}$, since at least $2t_2-2\in \mathcal O'$. 
To see this, note that we are considering $V>V_{c1}$, and hence $\log{V} > \log{V_{c1}} = G(2t_2-2)$, leading to $2t_2-2 \in \mathbb{R}$. 
We further note that $G(E)\to +\infty$ as $\abs{E}\to +\infty$. 
Therefore, due to the continuity of $G(E)$, for any real number $E\in \mathcal O'$, there always exists a complex number $E'$ that satisfies the following conditions 
\begin{align*}
    \Re(E')=E\ \&\ G(E')=\log{V}. 
\end{align*}
Consequently, we obtain $\mathcal O = \mathcal O'$.
Therefore, the real localized spectrum is just the boundary of the open set $\mathcal O'$, while every point inside $\mathcal O'$ corresponds to at least one complex localized state, which proves our statement.

Besides, from the above analysis, we know that if $E$ is an eigenvalue, then the statement that the state $\ket{E}$ is extended is equivalent to $G(\mathrm{Re}(E))>\log{V}$. 
Similarly, the statement that $\ket{E}$ is localized is equivalent to the condition that $G(\mathrm{Re}(E))\leq\log{V}$. 

\subsection[t2 transition]{The $t_2$ transition}

\begin{figure}[!]
\includegraphics[width=\columnwidth]{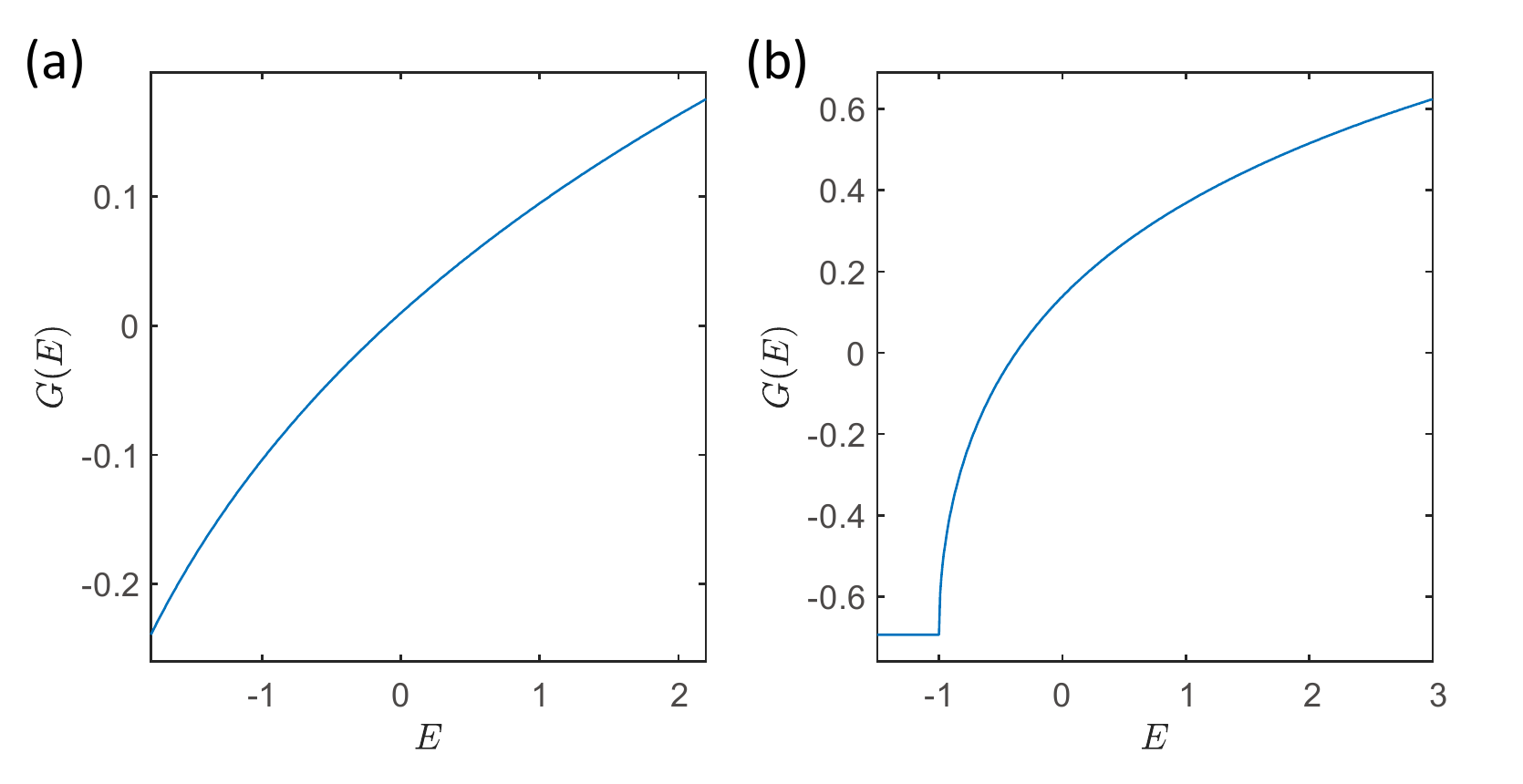}
\caption{\label{FigSM:GE} (a) and (b) plot $G(E)$ in Eq.~\eqref{EqSM:G(E)} with $t_2=0.1$ and $t_2=0.5$ on $U_E$, respectively. 
The set of energies on these two curves satisfying $G(E)>\log{V}$ belongs to extended states. }
\end{figure}

Additionally, another property of $G(E)$ is that $G(E)$ is a constant function on the interval $[-1/(4t_2)-2t_2,2]$. 
which implies that when $t_2\geq1/4$ there are a bunch of $E$ reaching the minimum of $G(E)$. 
As a result, multiple eigenstates will be localized simultaneously as $\log(V)$ reaches $V_{c1}$. 
This property explains the difference between panels (b)-(c) in Fig.~2 in the main text. 
One such example is shown in Fig.~\ref{FigSM:GE}, which plots $G(E)$ on $U_E$ for two different values of $t_2$. 
In particular, for a given $V$, all points on the curve satisfying $G(E)>\log{V}$ correspond to extended states in the system. 
Specifically, when $0 < t_2 <  1/4$ the $G(E)\cap U_E$ is a monotonic curve, and hence the number of localized states increases continuously from zero at $V = V_{c1}$. 
In contrast, when $ t_2 > 1/4$ the $G(E)\cap U_E$ has a plateau. 
As a result, the number of localized states is already finite at $V = V_{c1}$.

\subsection{The IPR and NPR in this model}

\begin{figure}[!]
\includegraphics[width=0.9\columnwidth]{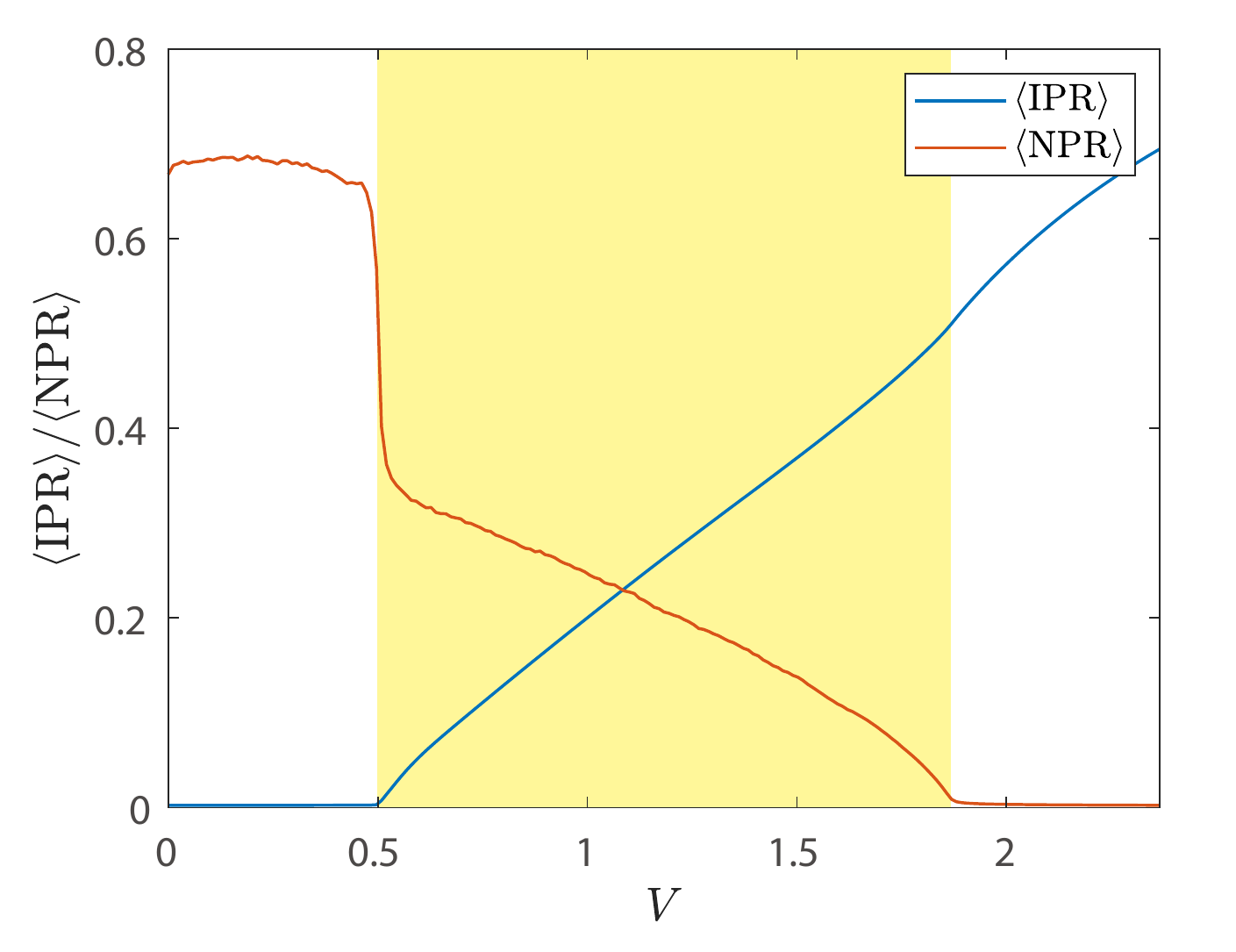}
\caption{\label{FigSM:IPR} 
In this figure we plot the $\mIPR$ and $\mNPR$ for the model in Eq.~\eqref{EqSM:TheModel} with $t_2=1/2$. 
The yellow background highlights the region when both extended and localized states appear in the energy spectrum. }
\end{figure}

One of the most widely used tools to diagnose the existence of ME is to plot the averaged inverse participation ratio (IPR) and normalized participation ratio (NPR) of all eigenstates~\cite{Li2017_PRB,Li2020_PRB}. 
In particular, these two quantities are defined as 
\begin{equation*}
    \text{IPR}^{(i)} = \sum_{n} \abs{u_{n}^{(i)}}^4, \quad 
    \text{NPR}^{(i)} = \qty[L\sum_{n}\abs{u_n^{(i)}}^4]^{-1}. 
\end{equation*}
In the above equation $L$ is the size of the one-dimensional system, the index $i$ labels different eigenstates, while the index $n$ labels different lattice sites. 
We will use $\mIPR$ and $\mNPR$ to denote the averaged IPR and NPR over all single-particle eigenstates, respectively. 

It is known that when the spectrum contains only extended states, we have $\mIPR\sim L^{-1}$ while $\mNPR$ is finite. 
In contrast, when the spectrum contains only localized states we have $\mNPR\sim L^{-1}$ while $\mIPR$ is finite. 
Only when the spectrum contains both extended and localized states, both $\mIPR$ and $\mNPR$ are finite. 

A plot of $\mIPR$ and $\mNPR$ for $t_2=1/2$ is shown in Fig.~\ref{FigSM:IPR}. 
We can see that the intermediate phase indeed appears for $V_{c1}<V<V_{c2}$, where $V_{c1} = 1/2$ and $V_{c2}=1+\sqrt{3}/2$ for $t_2=1/2$. 
It is interesting to note that the $\mIPR$ is almost constant in the extended phase ($0<V<V_{c1}$) for this non-Hermitian model, indicating that the localization length of the eigenstates does not decrease much when $V$ is increased from $0$ to $V_{c1}$. 
In contrast, for a Hermitian quasiperiodic model there is a clear decrease of $\mIPR$ in the extended phase as $V$ increases from $0$ to $V_{c1}$~\cite{Li2017_PRB,Li2020_PRB}. 
However, currently we do not fully understand the reason behind this difference.

\section{Further details on the experimental protocol}
In this section we provide additional details on the experimental protocol in Fig.~4 in the main text.

\begin{figure}[b]
\includegraphics[width=\columnwidth]{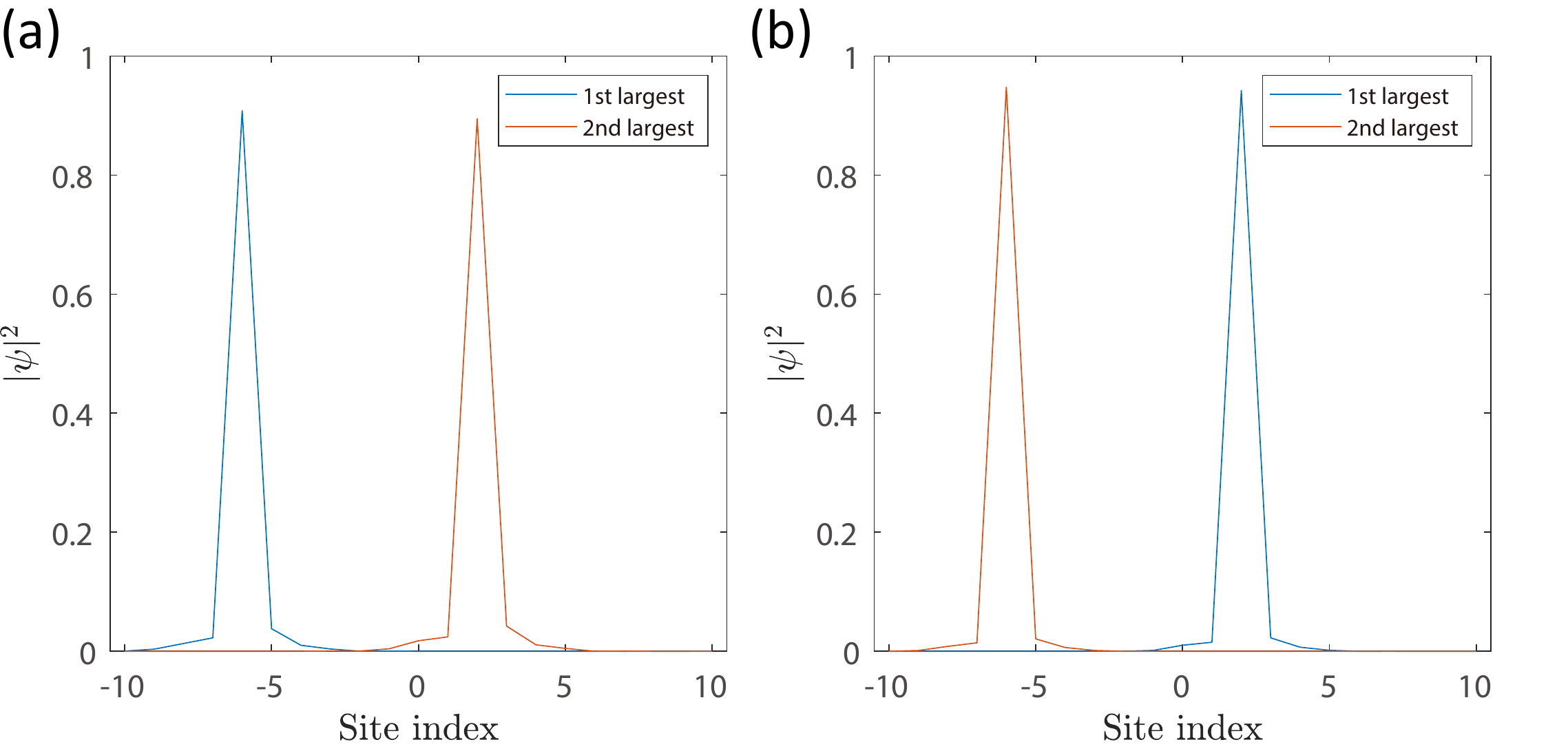}
\caption{\label{FigSM:Eigenstate} The two eigenstates with the largest imaginary energies in an $L=21$ system for the model in Eq.~\eqref{EqSM:TheModel}.  
In particular, the blue line corresponds to the state $\ket{\Emax}$ defined in the text. 
(a) and (b) show the results with $V=V_{c2}+1$ and $V=V_{c2}+2$, respectively. 
}
\end{figure}

\subsection[Long time]{The long-time dynamics}
We first focus on the long-time behavior of the system, and discuss the switching behavior observed in Fig.~4 in the main text. 

We first discuss panel (e) of Fig.~4 in the main text, which corresponds to $V = V_{c2} + 1$. 
In this case the state $\ket{\Emax}$ is localized around the $j=-6$ waveguide. 
Meanwhile, the eigenvalue of the state localized around the $j=2$ waveguide has an imaginary part that is only slightly smaller. 
Such a result can be seen in Fig.~\ref{FigSM:Eigenstate}(a). 
Consequently, the waveguide at $j=2$ quickly takes over the initial excitation, because it is much closer to the initially excited waveguide (at $j=0$) than the waveguide at $j$~$=$~$-6$. 
However, the state $\ket{\Emax}$ eventually dominates the dynamics at long times, and the excitation finally collapses onto the waveguide at $j=-6$. 

In contrast, when $V = V_{c2} + 2$, which corresponds to panel (f) of Fig.~4, the state $\ket{\Emax}$ now resides in the waveguide at $j=2$, as shown in Fig.~\ref{FigSM:Eigenstate}(b). 
As a result, the initial excitation quickly collapses onto the waveguide at $j=2$, and will not switch to other waveguides afterwards. 
It is also worth noting that the strong localization limit of $V = V_{c2} + 2$ can be understood directly from the potential distribution in Fig.~4(b) in the main text, because all eigenstates are well approximated by the Wannier functions in this limit, and the eigenvalues are also close to the potential energies in each waveguide. 
As we can see from Fig.~4(b) in the main text, $\Im(V_{j=2})$ is indeed the largest among all, while $\Im(V_{j=-6})$ is a close second. 
This is the fundamental reason why the dynamics in the strong localization limit is dominated by these two waveguides. 

\subsection{The short-time dynamics}

\begin{figure}[!]
\includegraphics[width=\columnwidth]{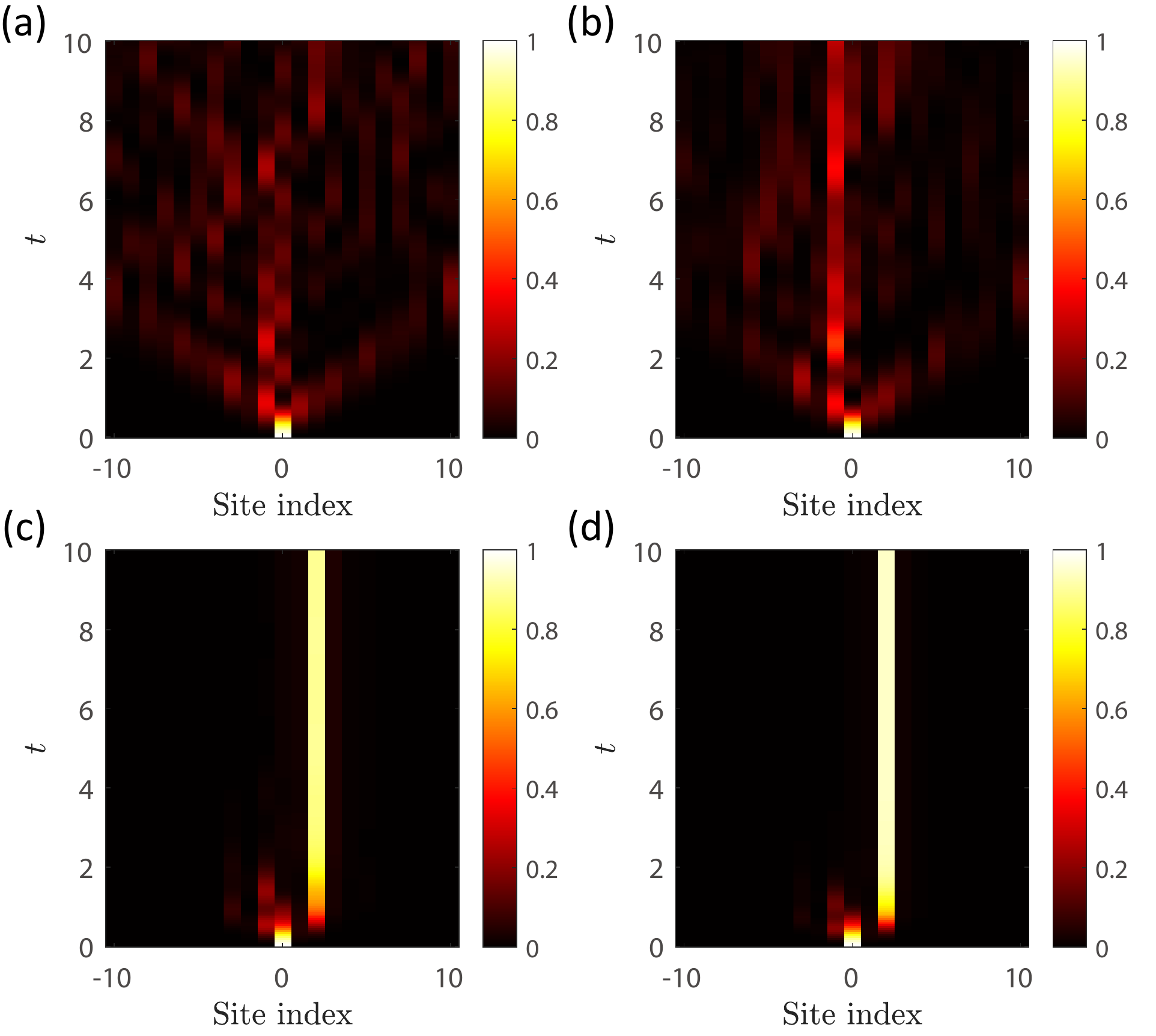}
\caption{\label{FigSM:Transmission} Short-time behavior of the experimental protocol shown in Fig. 4(a) in the main text. 
The system consists of $L=21$ waveguides, which are described by the model in Eq.~\eqref{EqSM:TheModel}. 
(a)-(d) show the quench dynamics starting from an initial excitation in the middle $j=0$ waveguide for $V=V_{c1}-0.1$, $V_{c1}+0.3$, $V_{c2}+1$, and $V_{c2}+2$, respectively. 
The unit of time is $\tau = \hbar/t_1$ in all figures. 
}
\end{figure}

Having understood the long-time dynamics, we now discuss the short-time behavior of the quench dynamics, which is shown in Fig.~\ref{FigSM:Transmission}. 
First, we observe that the initial excitation at $j=0$ was only retained in the system for less than $t=\tau_0$ in all four cases, where $\tau_0 = \hbar/t_1$ is a natural unit of time in our model. 
Second, in the presence of extended states ($V<V_{c2}$), the signal quickly spreads out to all waveguides. 
In particular, a light-cone-like structure is clearly visible, as shown in Fig.~\ref{FigSM:Transmission} (a)-(b). 
In contrast, when the spectrum contains localized states only ($V>V_{c2}$), the light-cone-like structure is gone. 
Instead, the initial excitation switches between different waveguides during the time evolution, and the exact dynamical process depends on the structure of the energy spectrum. 
For example, in Fig.~\ref{FigSM:Transmission}(c), when $\tau<t<2\tau$ the light propagates in the form of a state with appreciable amplitudes on the $j=-3$, $j=-1$, and $j=2$ waveguides. 
However, when $t>2\tau$ the amplitudes in the $j=-3$ and $j=-1$ waveguides gradually die off, and only the $j=2$ waveguide has an appreciable amplitude. 
The behavior in Fig.~\ref{FigSM:Transmission}(d) is similar to that in (c), although the amplitude in the $j=2$ waveguide is much stronger when $t>2\tau$.

\section{The case of complex hopping parameters}

In order to illustrate the versatility of our method, in this section we show that we can derive the exact ME condition even when both $t_1$ and $t_2$ are complex. 
This extension also has direct experimental applications because in photonic lattices it is possible to generate complex hoppings between neighboring waveguides. 
Specifically, we consider the following Hamiltonian, 
\begin{equation*}
    H=\sum_{j}\qty(t_1 e^{i\phi_1}c^{\dagger}_{j}c_{j+1}+t_2e^{i\phi_2}c^{\dagger}_{j}c_{j+2} + \text{h.c.}) 
    +\sum_j V_j n_{j}. 
\end{equation*}
In the above equation, the two phases can be reduced to one independent parameter $\phi_2-2\phi_1$ by the substitution $c_j^\dag\to e^{-i\phi_1j}c_j^\dag$. 
As a result, we shall consider the following Hamiltonian instead 
\begin{align}
    H = \sum_{ j}\qty(c^{\dagger}_{j}c_{j+1}+\tilde t_2 c^{\dagger}_{j}c_{j+2}+ V_j n_{j} + \text{h.c.}), 
    \label{EqSM:ComplexModel}
\end{align}
where we have defined $\tilde t_2=t_2e^{i\phi}$ with $\phi=\phi_2-2\phi_1$.
Note that we again set $t_1 = 1$ and $t_2>0$, in accordance with our convention in the main text. 

To proceed, we now apply the Fourier transformation in Eq.~\eqref{EqSM:FourierTransform} to rewrite Eq.~\eqref{EqSM:ComplexModel} in terms of $f(\theta)$ as 
\begin{align}
V f(\theta+\alpha)=\qty[E-2\cos\theta-2t_2\cos(2\theta-\phi)]f(\theta). \notag
\end{align}
Correspondingly, the function $G(E)$ and the set $U_E$ are defined respectively as
\begin{align}
    G(E) &= \dfrac1{2\pi}\int_{0}^{2\pi}\log\abs\big{E-g(\theta)}d\theta, \label{EqSM:G(E)_Complex}\\
    U_E &= \qty{E: \exists\, \theta, \text{s.t.} \; E=g(\theta)},
\end{align}
where $g(\theta) \equiv 2\cos\theta+2t_2\cos(2\theta-\phi)$.

\subsection[Exact expression for G(E)]{An exact expression for $G(E)$}

\begin{figure}[!]
\includegraphics[width=\columnwidth]{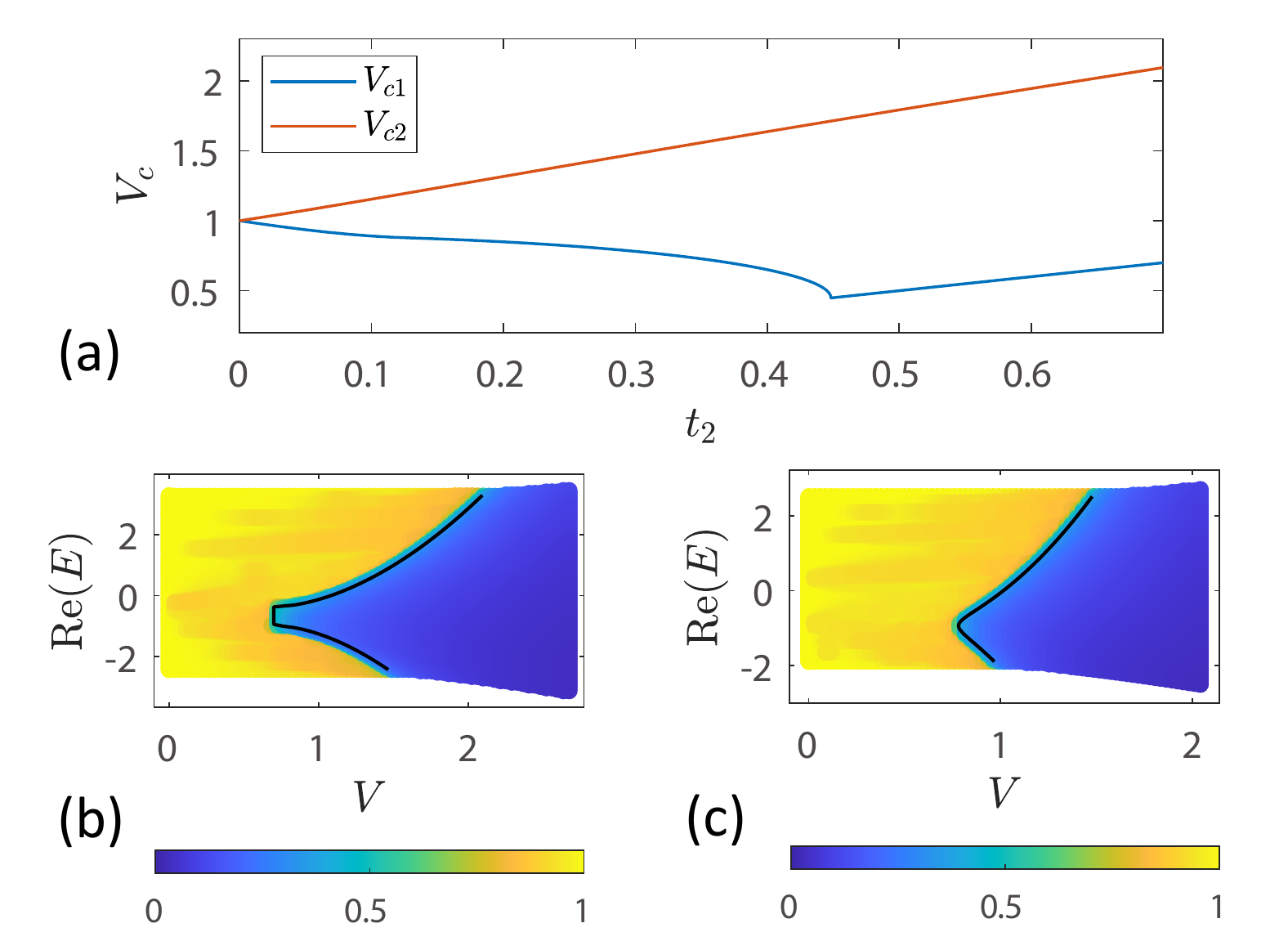}
\caption{\label{FigSM:ComplexHopping}(a) plots the two transition points $V_{c1},V_{c2}$ with respect to $t_2$ obtained numerically.  (b) and (c) plot $t_2=0.7$ and $t_2=0.3$ respectively. 
The color of (b) and (c) represents the fractal dimension of the eigenstates and the black lines are derived numerically from $G(E)=\log|V|$. 
Finally, $\phi_2-2\phi_1 = \pi/4$ in all three figures and the system size is $L=610$. 
}
\end{figure}

For the present model, we can still derive an analytic expression for $G(E)$: 
\begin{align}
    G(E)&=\frac1{2\pi}\int_{0}^{2\pi}\log|E-2\cos\theta-2t_2\cos(2\theta-\phi)|d\theta \notag\\
    &=\frac1{2\pi}\oint_{|z|=1}\log|z+1/z+\tilde t_2z^2+\tilde t_2^*/z^2-E|\frac{dz}{iz}\notag \\
    &=\frac1{2\pi}\oint_{|z|=1}\log|\tilde t_2z^4+z^3-Ez^2+z+\tilde t_2^*|\frac{dz}{iz} \notag \\
    &=\log{t_2}+\sum_{k=1}^4\frac1{2\pi}\oint_{|z|=1}\log|z-z_k|\frac{dz}{iz},
\end{align}
where $z_k$ are the four roots of the following equation: 
\begin{equation}
    \tilde t_2z^4+z^3-Ez^2+z+\tilde t_2^\ast=0. \label{EqSM:Roots} 
\end{equation}
If we further note that  
\begin{align}
\frac1{2\pi}\oint_{|z|=1}\log|z-z'|\frac{dz}{iz} =
\begin{cases}
    \log|z'|, & |z'|\geq1\\
    0,  & |z'|<1
\end{cases}, 
\end{align}
we find that 
\begin{align}
    e^{G(E)}=t_2\prod_{k=1}^{4}\max(|z_k|,1), \label{EqSM:ME-ComplexModel}
\end{align}
which satisfies $e^{G(E)}\geq t_2$. 
Because $z_k$ are just the roots of a quartic equation, they can still be written in closed forms. 
Besides, when $\phi=0$, we can substantially simplify $z_k$ by the following relation
\begin{align}
    0&=z+1/z+t_2z^2+t_2/z^2-E \notag\\ 
    &=(z+1/z)+t_2(z+1/z)^2-(E+2t_2),
\end{align}
which can be used to derive Eq.~\eqref{G(E)_real}.

In Fig.~\ref{FigSM:ComplexHopping} (b)-(c) we compare the predicted ME in Eq.~\eqref{EqSM:ME-ComplexModel} with the single-particle spectrum in a finite system. 
In particular, we choose $\phi = \pi/4$, and study two cases with $t_2 = 0.7$ and $t_2 = 0.3$, respectively. 
We find that our theoretical predictions of the ME (black solid line) agrees well with the numerical results.

\subsection[An exact expression for tc]{An exact expression for $t_2^{(c)}$}
Another interesting observation from Fig.~\ref{FigSM:ComplexHopping}(a) is that the structure of the ME is again different for small $t_2$ and large $t_2$, although the critical point now shifts to $t_2\approx 0.45$ instead of $1/4$ for the $\phi=0$ case.
We now show that this critical $t_2^{(c)}$ can again be determined analytically. 
This can be done without writing down $z_k$ explicitly. 

To begin with, note from Eq.~\eqref{EqSM:Roots} that we have $z_1z_2z_3z_4=\tilde t_2^*/\tilde t_2$, which implies that  
\begin{align}
    \sum_{k=1}^4\log|z_k|=\log|\tilde t_2^*/\tilde t_2|=0.
\end{align}
As a result, we note from Eq.~\eqref{EqSM:ME-ComplexModel} that  
\begin{align}
    G(E)&=\sum_{k=1}^4\max\left(\log|z_k|,0\right)+\log t_2\\
    &=\frac12\sum_{k=1}^4\abs\bigg{\log|z_k|}+\log t_2\geq \log t_2.
\end{align}
Hence, the last equality holds if and only if all four roots satisfy $|z_k|=1$. 

We know that $V_{c1}=\min e^{G(E)}$, so if there exists an energy $E$ that allows such four roots, then $V_{c1}=t_2$, independent of $\phi$. 
Besides, if there are more than one such $E$, they all turn into localized states at once as $V$ passes $t_2$. 
Hence, to determine the critical $t_2$, we need to check whether Eq.~\eqref{EqSM:Roots} allows four roots on $|z|=1$ for some $E$. 
Equivalently, we can ask whether the equation $E=2\cos\theta+2t_2\cos(2\theta-\phi)$ allows four roots in a period for some $E$. 
This condition amounts to ask whether $2\sin\theta+2t_2\sin(2\theta-\phi)=0$ has four roots. 
Based on this observation, we can derive the expression for the critical point $t_2^{(c)}$ as follows, 
\begin{equation}
    t_2^{(c)}(\phi)=\frac{1}4\left[\left|\sin \frac\phi2\right|^{2/3}+\left|\cos \frac\phi2\right|^{2/3}\right]^{3/2}. 
\end{equation}
In particular, when $\phi=\pi/4$ we have $t_2^{(c)} \approx 0.448$, in agreement with the observation in Fig.~\ref{FigSM:ComplexHopping}(a).

\bibliographystyle{apsrev4-2}
\bibliography{t1t2Model_v6.bib}
